\documentclass[aps,pra,superscriptaddress,twocolumn]{revtex4-2}

\usepackage{graphicx} 
\usepackage{amsmath,amssymb,xcolor,bm}
\usepackage{comment}
\usepackage{upgreek}
\usepackage{xfrac}
\usepackage{braket}
\usepackage{multirow}
\usepackage{bbold}
\usepackage{url}
\usepackage{mathtools}
\usepackage{xcolor}
\usepackage{enumerate}
\usepackage{ulem}

\definecolor{darkgreen}{RGB}{1,150,62}
\definecolor{orange}{RGB}{235, 129, 0}
\usepackage[bookmarks=true,
            bookmarksnumbered=false,
            bookmarksopen=true,
            breaklinks=true,
            pdfborder={0 0 0},
            final=true,
            backref=none,
            colorlinks=true,
            linkcolor=blue,
            menucolor=blue,
            citecolor=blue,
            urlcolor=blue]{hyperref}

\begin{document}

\title{Benchmarking Quantum Convolutional Neural Networks and their classical surrogates on long-range strongly correlated models}

\author{Tatiana A. Bespalova}
\affiliation{Universit\'e de Strasbourg and CNRS, ISIS (UMR 7006), 67000 Strasbourg, France}

\author{Réka Schwengelbeck}
\affiliation{Universit\'e de Strasbourg and CNRS, ISIS (UMR 7006), 67000 Strasbourg, France}

\author{Laura Mismetti}
\affiliation{IBM Research, S\"aumerstrasse 4, 8803 R\"uschlikon, Switzerland}

\author{Johannes Schachenmayer}
\affiliation{Universit\'e de Strasbourg and CNRS, ISIS (UMR 7006), 67000 Strasbourg, France}

\author{Ivano Tavernelli}
\affiliation{IBM Research, S\"aumerstrasse 4, 8803 R\"uschlikon, Switzerland}

\author{Francesco Tacchino}
\affiliation{IBM Research, S\"aumerstrasse 4, 8803 R\"uschlikon, Switzerland}

\date{\today}

\begin{abstract}
    We benchmark quantum convolutional neural networks (QCNNs) and their classical surrogates for phase differentiation in one-dimensional strongly correlated systems with beyond-nearest-neighbour terms. Using an anisotropic Heisenberg spin chain and an extended $t$-$t'$ Fermi-Hubbard model as representative spin and fermionic testbeds, we train QCNN classifiers on labelled ground-state datasets and map out finite-size phase diagrams. 
    In hybrid setups, we employ Pauli backpropagation to express the QCNN measurement as a weighted sum of Pauli-string expectation values to be evaluated directly on input states, and study how truncation parameters control the trade-off between approximation accuracy and the size of the resulting Pauli expansion. 
    Robust convergence is achieved for the Heisenberg model under aggressive truncation, whereas the Fermi-Hubbard model requires substantially looser thresholds, resulting in a sharp growth in retained Pauli strings and evaluation cost.
    Our results contribute to charting where hybrid quantum-classical training is viable for learning strongly correlated quantum states, and provide practical guidance for choosing truncation regimes in Pauli backpropagation-based surrogates.
\end{abstract}

\maketitle

\section{Introduction}

Identifying and characterizing quantum phases of matter from microscopic data is a central task in the study of strongly correlated many-body systems. Conventional approaches typically rely on physically motivated quantities like order parameters, excitation gaps, or correlation functions, whose selection depends on prior analysis and understanding of the underlying Hamiltonian~\cite{sachdev1999quantum}. 
An alternative, particularly appealing perspective is that the information that distinguishes phases is encoded in the many-body ground state itself through its pattern of quantum correlations. From this viewpoint, phase differentiation can be framed as a learning problem in which the input data are quantum states and the goal is to deduce phase structure without explicitly constructing the relevant observables. 

This second point of view has motivated numerous studies on the application of machine learning to phase recognition, both in classical and quantum settings. Classical neural-network approaches have been shown to identify phases and phase transitions from microscopic data such as spin configurations or measurement samples, often without handcrafted features~\cite{carrasquilla2017machine, van2017learning, dong2019machine}. Quantum machine learning aims to extend this paradigm by processing quantum data directly, potentially accessing nonlocal features that may be difficult to capture with classical data.

Quantum convolutional neural networks (QCNNs)~\cite{cong2019quantum} were introduced as a quantum analogue of classical convolutional architectures, designed to process many-body quantum data through an alternation of locality-preserving layers and progressive ``pooling'', a step that reduces the number of active degrees of freedom, followed by the measurement of an output observable on the remaining effective dimensions. 
QCNNs have been applied to phase classification tasks in several quantum lattice models, with the promise that a parametrised quantum circuit can learn phase-relevant features directly from ground states~\cite{umeano2025can, sander2025quantum}.
At the same time, recent work has raised the question of which components of QCNN-based approaches genuinely require quantum processing. In particular, the local, structured and effectively shallow nature of common QCNN architectures, which renders them easily trainable~\cite{pesah2021absence},  can also make their evaluation well-suited to efficient classical approximation in certain regimes~\cite{bermejo2024quantum}.  
One prominent route to such approximations is to propagate the output observable backwards through the circuit in the Heisenberg picture, yielding a Pauli-string expansion whose truncation controls an explicit accuracy-cost trade-off~\cite{fontana2025classical, rudolph2025pauli}. 
The method is particularly effective when input states are classically easy to manipulate~\cite{rudolph2023classicalsurrogatesim}. Crucially, under specific locality of correlations assumptions, these techniques were recently employed to reduce QCNNs to a classical surrogate to be evaluated via suitable measurements (e.g., classical shadows) directly on the input states, hence bypassing coherent quantum execution~\cite{bermejo2024quantum}. 

In this work, we investigate the practicality of this approach in cases where the dataset consists of potentially strongly correlated states, comparing truncated classical surrogates with full QCNN forward execution. More specifically, we investigate phase differentiation with QCNNs and their classical surrogates for strongly correlated one-dimensional models with terms beyond nearest neighbours, assessing both computational and perspective sampling costs, as well as the overall classification performances. 
In our numerics, we treat ground states as quantum inputs, obtained by exact diagonalisation, and focus exclusively on the subsequent analysis part.  
In an actual experimental workflow, these states could be prepared via a certain quantum process (e.g., adiabatic state preparation or a variational ground-state optimisation), to then either be supplied to a coherent QCNN, or be used to extract the expectation values of Pauli-string observables required in a Heisenberg-picture surrogate. 
To compare the quantum and classical-surrogate versions of the training we first quantify the endpoint fidelity of truncated Pauli backpropagation at the parameters optimised with forward QCNN circuit evaluation and measure how the approximation error and the number of retained Pauli strings depend on truncation thresholds. Second, we perform end-to-end optimisation runs in which QCNN training is carried out using a truncated Pauli-backpropagation surrogate at every step, thereby directly testing surrogate trainability rather than only endpoint agreement.

As testbeds, we choose an extended Fermi-Hubbard model with next-to-nearest-neighbour hopping and an anisotropic Heisenberg model with next-to-nearest-neighbour interactions. 
Both exhibit nontrivial many-body correlations and a rich phase structure, while their phase diagrams are sufficiently well understood to provide reliable benchmarks.
{Comparing the spin and fermionic models reveals a pronounced model dependence: in the Heisenberg case, surrogate training can succeed already under aggressive truncation, whereas in the fermionic Hubbard case the truncation required for reliable training becomes substantially more demanding, consistent with the non-local operator structure induced by fermion-to-qubit mappings such as Jordan-Wigner.}

The remainder of the paper is organised as follows. In Sec.~\ref{sec:methods} we describe the QCNN architecture and training protocol, introduce Pauli backpropagation, and present the classical-surrogate optimisation procedure.
Sec.~\ref{sec:models} defines the extended Fermi-Hubbard and anisotropic Heisenberg models studied.
In Sec.~\ref{sec:results} we report phase-classification results for QCNN training, quantify truncation accuracy and cost for Pauli backpropagation, and compare direct QCNN training to surrogate optimisation. 
We conclude in Sec.~\ref{sec:conclusion} with a discussion of implications for hybrid quantum-classical phase differentiation and potential extensions to larger system sizes and alternative encodings.

\section{Methods}
\label{sec:methods}

This section summarizes the algorithmic procedures used throughout the paper. In Sec.~\ref{methods:qcnn} we introduce the QCNN architecture and the supervised training protocol used for phase differentiation from ground-state inputs. In Sec.~\ref{methods:backpropagation} we describe Pauli backpropagation, which rewrites the QCNN output in the Heisenberg picture as a truncated weighted sum of Pauli-string expectation values on the input state. Finally, Sec.~\ref{methods:fullyclassical} uses this backpropagation procedure to define a fully classical surrogate for QCNN optimisation.

\subsection{QCNN training for phase differentiation}
\label{methods:qcnn}

\begin{figure}[tb]
    \centering
    \includegraphics[width=1.0\linewidth]{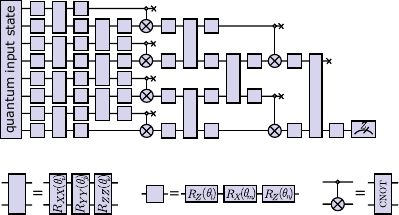}
    \caption{Example QCNN architecture for $N=8$ qubits. Parametrized two-qubit entangling gates and single-qubit rotations, each instance of a gate parametrized independently from the others, are interleaved with pooling layers that reduce the number of active qubits. Pooling can be implemented either by measuring and discarding selected qubits (ignoring the measurement outcomes) or by tracing them out; for our purposes these are equivalent descriptions. The final prediction is obtained from a single-qubit $Z$ measurement on the remaining output qubit.}
    \label{fig:circuit}
\end{figure}

\paragraph{Learning task and data.}
Our goal is supervised phase differentiation from many-body ground states. Rather than encoding Hamiltonian parameters, the QCNN operates directly on the corresponding ground state wavefunctions, reflecting the assumption that these states already contain the correlations required to distinguish different phases. 
We construct a labelled training set $\{(\ket{\psi_i}, \ell_i)\}_{i=1}^{N_{\mathrm{train}}}$ of ground states $\ket{\psi_i}$ together with class labels $\ell_i\in\{-1,+1\}$, and a separate test set used to evaluate generalisation performance. 
In this work, ground states are obtained by exact diagonalisation as a proof-of-principle. 

\paragraph{QCNN architecture.}
QCNNs are constructed as parametrized quantum circuits composed of repeated local ``convolution'' blocks and pooling operations, leading to a progressive reduction in the number of active qubits. In our implementation, pooling is realised by discarding a subset of qubits after applying an entangling layer; equivalently one may measure and discard these qubits while ignoring the measurement outcomes. 
The architecture employed throughout this work is illustrated in Fig.~\ref{fig:circuit}. Compared to the architecture of Ref.~\cite{nagano2023quantum}, we remove parametrized gates that do not increase the circuit expressivity (i.e., in our case, ability to differentiate between a wide variety of quantum states) for the present task, thereby reducing the number of trainable parameters (by approximately $35\%$ in our implementation) and improving optimisation efficiency. Further details on QCNN circuit structures are provided in Appendix~\ref{appendix:structureN12}.

\paragraph{Training objective and optimisation.}
Given circuit parameters $\boldsymbol{\theta}$, the QCNN produces a scalar prediction $f_{\psi_i}(\boldsymbol{\theta})$ for each input state $\ket{\psi_i}$, obtained from the expectation value of the final $Z$ measurement on the output qubit. We train the circuit by minimizing a mean-squared-error loss,
\begin{equation}
    \lambda(\boldsymbol{\theta})=\sum_{i=1}^{N_{\mathrm{train}}}\left[f_{\psi_i}(\boldsymbol{\theta})-\ell_i\right]^2.
\end{equation}
The optimisation is performed with a classical optimiser; in this work we use gradient descent with adaptive learning rate: if not stated otherwise, we start with the learning rate $\eta_0=100$, increasing it by $10\%$ after each successful step and decreasing by $33\%$ after a step that increases the loss. 
We initialize all circuit parameters to the fixed value $\theta=1$ for all gates to ensure reproducibility across runs, or use a uniform random initialisation $\theta \in [0, 2 \pi]$ when stated explicitly.

The resulting trained QCNN serves as the reference against which the Pauli-backpropagation-based surrogate methods introduced in the following sections are compared 
(e.g., with respect to classification performance, accuracy and/or confusion matrices; see Sec.~\ref{results:qcnn_training} and Appendix~\ref{appendix:confusion_matrices}).

\subsection{Pauli backpropagation of QCNN observables}
\label{methods:backpropagation}

\begin{figure}[tb]
    \centering
    \includegraphics[width=1.0\linewidth]{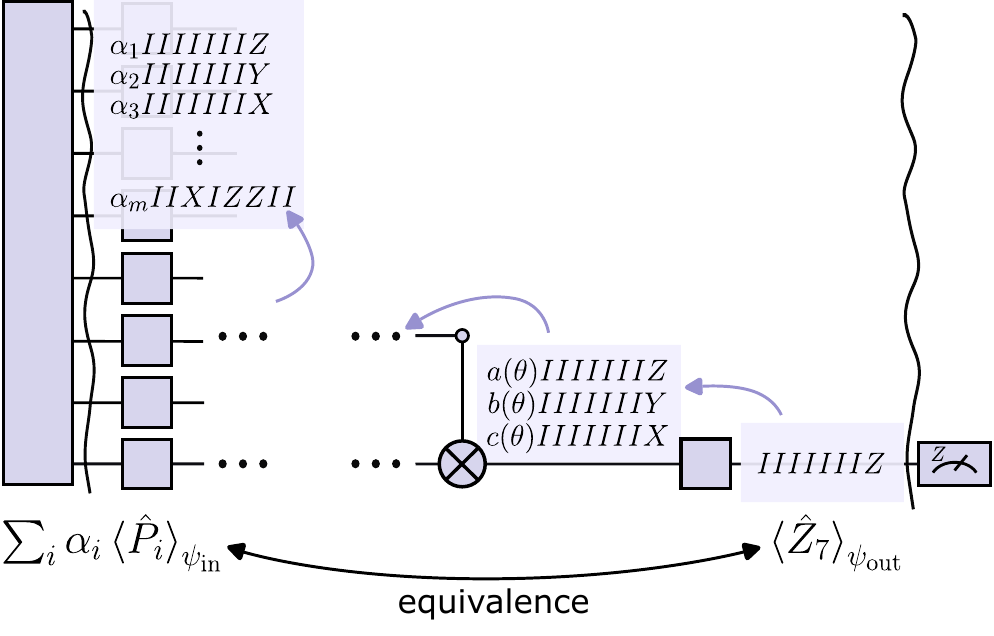}
    \caption{Illustration of Pauli backpropagation in a QCNN. The final measurement observable (e.g., $\hat Z_7$) is evolved backwards through the network 
    by successive unitary conjugation: after each unitary layer ($U_l$), it updates as ($O_{l-1}=U_l^\dagger O_l U_l$).
    The resulting effective observable acting on the input can then be written as a linear combination of Pauli strings $\sum_i \alpha_i \hat P_i $.
    }
    \label{fig:backpropagation}
\end{figure}

Our goal is to investigate to what extent the quantum processing performed by a QCNN can be replaced by a classical surrogate while preserving its ability to distinguish quantum phases. To this end, we employ Pauli backpropagation~\cite{fontana2025classical, rudolph2025pauli}, where we rewrite the QCNN output observable in the Heisenberg picture as an effective observable acting directly on the input state. Rather than executing the QCNN on quantum hardware, this representation enables the network output to be approximated classically through a truncated expansion in Pauli operators~\cite{bermejo2024quantum}.

Let us consider a general case of a parametrized circuit $\hat{U}(\boldsymbol{\theta})$ acting on an input state $\ket{\psi_\text{in}}$ and resulting in a new quantum state $\ket{\psi_\text{out}} = \hat{U}(\boldsymbol{\theta}) \ket{\psi_\text{in}}$. Instead of measuring a Pauli observable $\hat O$ on the resulting output state $\langle \hat{O} \rangle = \bra{\psi_{\mathrm{out}}}\hat{O}\ket{\psi_{\mathrm{out}}}$, Pauli backpropagation computes the corresponding Heisenberg-evolved operator 
\begin{equation}
\hat{O}_{\mathrm{in}}(\boldsymbol{\theta}) = \hat{U}^\dagger(\boldsymbol{\theta})\,\hat{O}\,\hat{U}(\boldsymbol{\theta}),
\end{equation}
and evaluates its expectation value directly on the input state,
$\langle \hat{O} \rangle = \bra{\psi_\text{in}}\hat{O}_{\mathrm{in}}(\boldsymbol{\theta})\ket{\psi_\text{in}}$. 
This backpropagation approach is sketched in Fig.~\ref{fig:backpropagation} for $N=8$.
As the Pauli operators form a complete basis for operators on the qubit Hilbert space, the propagated observable can always be expressed as a linear combination of Pauli strings. Clifford gates simply map Pauli strings onto other Pauli strings, whereas parametrized single-qubit rotations generate linear combinations of Pauli strings.
For example, for a single-qubit rotation $R(\boldsymbol{\theta})$ one can write $R^\dagger(\boldsymbol{\theta})\,P_i\,R(\boldsymbol{\theta})=\sum_j c_{ji}(\boldsymbol{\theta})P_j$ with $P_i\in\{I,X,Y,Z\}$.
Consequently, evaluating the QCNN reduces to computing expectation values of Pauli strings on the input state
\begin{equation}
\label{eq:backprop_decomp}
\langle Z_{q_\text{obs}} \rangle_{\psi_\text{out}} = \sum_{i=1}^{N_\text{Pauli strings}} \alpha_i(\boldsymbol{\theta})\,\langle P_i \rangle_{\psi_\text{in}}.
\end{equation}
So far, this representation is exact. However, the number of terms ($N_\text{Pauli strings}$) in the expansion in Eq.~\eqref{eq:backprop_decomp} generally grows rapidly with circuit depth and the number of non-Clifford operations, making an exact classical evaluation expensive.
{Indeed, if the circuit is expressive enough all $4^N$ Pauli strings (for a system of $N$ qubits) could appear in Eq.~\eqref{eq:backprop_decomp}, see also Table~\ref{tab:costs} in Section~\ref{results:surrogate_training}}.  To keep the classical cost manageable, the propagation is truncated in a controlled way using thresholds that limit which Pauli terms are retained~\cite{rudolph2023classicalsurrogatesim,schuster2024polynomialtimeclassicalalgorithmnoisy}. 
In this work, we employ three truncation schemes. First, we impose a maximum Pauli-weight cutoff (PW), retaining only Pauli strings acting non-trivially on at most a prescribed number of qubits. Second, we apply a maximum-frequency cutoff ($\omega)$, which removes high-frequency Fourier components in the parameter-dependent coefficients $\alpha_i(\boldsymbol{\theta})$ during propagation~\cite{rudolph2023classicalsurrogatesim}. Third, we apply a minimum absolute coefficient cutoff ($\varepsilon$), retaining only terms with coefficients of at least a specific magnitude $|\alpha_i|$. 
Thus the indices of $i$ from Eq.~\eqref{eq:backprop_decomp} only run through those of the coefficients that correspond to the criteria above resulting in a reduced number of terms:
\begin{equation}
\label{eq:backprop_decomp_trunc}
\langle Z_{q_\text{obs}} \rangle_{\psi_\text{out}} = \sum_{\tilde i=1}^{N_\text{Pauli strings}(\text{PW}, \omega, \varepsilon)} \alpha_{\tilde i}(\boldsymbol{\theta})\,\langle P_{\tilde i} \rangle_{\psi_\text{in}}.
\end{equation}
By systematically varying these truncation parameters, we investigate the trade-off between classical computational cost and the ability of the resulting surrogate to reproduce the predictions of the full QCNN.

\subsection{QCNN optimisation using classical surrogates}
\label{methods:fullyclassical}

Using the tools described in the previous section, we may replace the forward execution of a QCNN with the evaluation of its classical surrogate, while leaving the optimisation procedure unchanged.
In practice, at every optimisation step one first recomputes, given the current parameter vector $\boldsymbol{\theta}$, a truncated Pauli expansion and then evaluates the required expectation values on the input states to compute the loss~\cite{rudolph2023classicalsurrogatesim}. 

More explicitly, given a QCNN architecture (Fig.~\ref{fig:circuit}), a loss function, a classical optimiser, and an initial parameter vector $\boldsymbol{\theta}_0$, each optimisation step proceeds as follows: (i) backpropagate the final measurement observable through the QCNN at parameters $\boldsymbol{\theta}$, applying the chosen truncation rules (controlling PW, $\omega$, $\varepsilon$), to obtain a list of Pauli strings $\{P_i\}$ and coefficients $\{\alpha_i(\boldsymbol{\theta})\}$. (ii) For each training input $\ket{\psi_k}$, evaluate the expectation values $\langle P_i\rangle_k=\bra{\psi_k}P_i\ket{\psi_k}$ required by Eq.~\eqref{eq:backprop_decomp_trunc}. Note that only the Pauli coefficients $\alpha_i(\boldsymbol{\theta})$ depend on the trainable circuit parameters, whereas the expectation values $\langle P_i \rangle_{\psi_\text{in}}$ depend solely on the input states. Thus, for a fixed training data set, the Pauli expectation values can in principle be computed once and reused during the optimisation.  (iii) Compute the model outputs from Eq.~\eqref{eq:backprop_decomp_trunc} and evaluate the loss on the training set. (iv) Update $\boldsymbol{\theta}$ using the chosen classical optimiser and repeat until a stopping criterion is met (e.g., loss change below a threshold or a fixed iteration budget).

At the end of the optimisation, we evaluate the circuit with the trained parameters on the held-out test set to report classification performance. By varying the truncation parameters in Pauli backpropagation, this framework allows us to investigate the trade-off between the computational cost of the surrogate and its ability to reproduce the predicitions of the QCNN.

In case when the input states themselves are given as quantum circuits, one could in principle extend this procedure by backpropagating, e.g., a variational ground-state preparation ansatz, together with the QCNN. 
However, for strongly correlated systems at large scale, it is reasonable to expect that this would substantially enlarge the resulting Pauli expansion and become quickly computationally intractable.
In this work, we restrict surrogate backpropagation to the QCNN part only.

\section{Models}
\label{sec:models}
We benchmark model performances on two strongly correlated open-boundary chain models with hopping or interaction terms beyond nearest neighbours. Specifically, we choose an anisotropic Heisenberg model as a representative of spin systems and a Fermi-Hubbard model as a representative of fermionic systems. 
In the following subsections, we specify the exact configurations of the models and briefly discuss the physics in the chosen parameter regimes.

\begin{figure}[tb]
    \centering
    \includegraphics[width=0.85\linewidth]{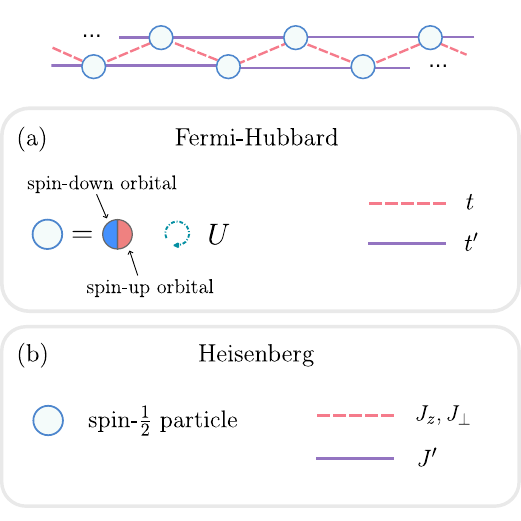}
    \caption{
    The structure of an open one-dimensional chain with nearest and next-to-nearest neighbours connectivity. (a) In the Fermi-Hubbard model, each site corresponds to two spin-orbitals (up and down), $U$ is the coefficient of onsite interaction, $t$ and $t'$ are nearest and next-to-nearest hopping coefficients respectively. 
    (b) In the Heisenberg model, each site corresponds to one spin-$\frac{1}{2}$  particle, 
    $J_z$ and $J_\perp$ are the coefficients of anisotropic nearest-neighbour exchange interactions in the longitudinal and transverse directions, and $J'$ is the coefficient of the strength of the isotropic next-to-nearest neighbour coupling. }
    \label{fig:models}
\end{figure}

\subsection{Fermi-Hubbard model with next-to-nearest neighbour hopping}

We consider the one-dimensional extended Fermi-Hubbard model with nearest- and next-to-nearest-neighbour hopping under open boundary conditions, as finite-size representation of the thermodynamic-limit (see Fig.~\ref{fig:models}). The inclusion of next-to-nearest-neighbour hopping captures longer-range electron delocalisation and introduces kinetic frustration through competing hopping processes, leading to a richer phase diagram than in the standard one-dimensional Hubbard model~\cite{phan2025ab,boman1998identification,raghu2002structural,delannoy2009low}. The resulting increase in the complexity of the ground-state structure provides a demanding benchmark for phase differentiation and classical surrogate models. In the limit of vanishing next-to-nearest-neighbour hopping, the model reduces to the Hubbard model~\cite{Hubbard1963,Gutzwiller1963,Kanamori1963}. 

The Hamiltonian for a chain of length $L$ is
\begin{equation}
\begin{split} \label{eq:HamiltonianFH}
    \hat H = & - \Big[\sum\limits_{\sigma, j=1}^{{L-1}}
    t \, \hat c_{j,\sigma}^\dagger \hat c_{j + 1, \sigma}  +
    \sum\limits_{\sigma, j=1}^{{L-2}} t' \, \hat c_{j,\sigma}^\dagger \hat  c_{j + 2, \sigma} + \Big. \\  &\Big.  + \text{H.c.} \Big] +U\sum\limits_{j = 1}^{{L}} \hat  n_{j, \uparrow} \hat n_{j, \downarrow} \text,
\end{split}
\end{equation}
where $\hat c_{j,\sigma}^\dagger$ ($\hat c_{j,\sigma}$) creates (annihilates) a fermion with spin $\sigma\in{\uparrow,\downarrow}$ at lattice site $j$, and $\hat n_{j,\sigma}=\hat c_{j,\sigma}^\dagger\hat c_{j,\sigma}$ is the corresponding number operator. The last interaction term describes onsite repulsion between fermions of opposite spin with strength $U$, while the hopping terms account for particle tunneling between nearest and next-to-nearest neighbouring sites with amplitudes $t$ and $t'$, respectively. 

In this work, we restrict our analysis to repulsive interactions ($U>0$), positive hopping amplitudes and on half-filled systems without spin-imbalance. We use $t\equiv 1$ as the unit of energy and, unless stated otherwise, study finite chains of $L=4$ and $L=6$ sites. In the thermodynamic limit, the half-filled $t$-$t'$ Hubbard model exhibits three main regimes. For $t'/t<1/2$, the ground state is a Mott insulator,  with gapless spin excitations (C0S1), where C$n$S$m$ denotes a phase with $n$ gapless charge modes and $m$ gapless spin modes~\cite{balents1996weak,lieb1968absence}. For $t'/t>1/2$, increasing frustration leads either to a fully gapped spontaneously dimerized insulating phase (C0S0) at sufficiently large $U/t$, or to a spin-gapped metallic phase (C1S0) with dominant superconducting correlations at smaller $U/t$~\cite{fabrizio1996superconductivity}.
Throughout this work, we compare our finite-size results with these established phase boundaries.

We employ the Jordan-Wigner transformation~\cite{jordan1928paulische} to map the fermionic Hamiltonian into a qubit model. We adopt a block encoding in which the first $L$ qubits correspond to spin-up orbitals and the remaining $L$ qubits correspond to spin-down orbitals.
Ground states sampled across the $(U/t,t'/t)$ parameter space are encoded in the qubit basis and provide the dataset used for QCNN training and accuracy evaluation.

\subsection{Anisotropic Heisenberg model with next-to-nearest neighbour interaction}

To complement the fermionic case, we investigate the anisotropic Heisenberg spin chain, which serves as a standard model for quantum magnetism. Following the sign conventions established in Ref.~\cite{PlekhanovB2010phase}, we extend the standard model to include next-to-nearest neighbour couplings. The Hamiltonian under open boundary conditions is then given by

\begin{equation}
\begin{split} \label{eq:HamiltonianHeis}
    \hat H = -J_z \sum_{i=1}^{L-1} \hat S^z_i \hat S^z_{i+1} + J_\perp \sum_{i=1}^{L-1} \left( \hat S^x_i \hat S^x_{i+1} + \hat S^y_i \hat S^y_{i+1}\right) \\ + J' \sum_{i=1}^{L-2} \mathbf{\hat S}_i \cdot \mathbf{\hat S}_{i+2}, 
\end{split}
\end{equation}
where ${\hat S}^\alpha_i$ ($\alpha \in \{ x, y, z\})$ are the standard spin-1/2 operators at site $i$, and $\mathbf{\hat S_i} = \left(\hat S^x_i, \hat S^y_i, \hat S^z_i\right)$ is the vector spin operator. The coefficients 
$J_z$ and $J_\perp$ characterise the anisotropic nearest-neighbour exchange interactions in the longitudinal and transverse directions, respectively. More precisely, $J_z$ controls the coupling between the $z$-components of neighbouring spins, and $J_{\perp}$ governs the transverse exchange terms $\hat S^+_i \hat S^-_{i+1} + \hat S^-_i \hat S^+_{i+1} $, generating quantum fluctuations in the $xy$-plane. The parameter $J'$ corresponds to the strength of the isotropic next-to-nearest neighbour coupling.
The competition between the anisotropic nearest neighbour couplings ($J_z, J_{\perp}$) and the isotropic next-to-nearest-neighbour interaction $J'$ gives rise to a rich zero-temperature phase diagram, which has been extensively investigated in Refs.~\cite{PlekhanovA2006, PlekhanovB2010phase}. 
We list the principal phases below. 

\textit{Ferromagnetic (FM):} $0<J_\perp<J_z$ and $J'\le0.31J_z$. Long-range ferromagnetic order.
    
\textit{Spin-fluid I (SF-I):} $J_\perp>J_z$. Gapless XXZ-like phase with quasi-long-range antiferromagnetic correlations and dominant wave vector $q=\pi$.

\textit{E-I:} $0<J_\perp<J_z$ and $0.31J_z<J'<0.43J_z$. Exponentially decaying spin correlations with residual ferromagnetic character.

\textit{E-II:} Between the two spin-fluid phases. Exponentially decaying correlations arising from the competition between nearest- and next-to-nearest-neighbour interactions.

\textit{Spin-fluid II (SF-II):} $J'>0.43J_z$. Gapless phase with quasi-long-range correlations and dominant wave vector $q=\pi/2$.

Throughout this work, these thermodynamic-limit phase boundaries serve as a reference for interpreting our finite-size phase differentiation results.

In contrast to the Fermi-Hubbard model, each Heisenberg site $i$ directly maps to a single qubit, 
and each spin operator $\hat{S}^\alpha_i$ maps directly onto a single-qubit Pauli operator 
acting on qubit $i$ via
\begin{equation}
    \hat{S}^\alpha_i = \frac{1}{2}\, \hat{\mathcal{P}}^\alpha_i, \qquad
    \hat{\mathcal{P}}^\alpha_i \equiv 
    \underbrace{I \otimes \cdots \otimes I}_{i-1} 
    \otimes\, \sigma^\alpha \,\otimes 
    \underbrace{I \otimes \cdots \otimes I}_{L-i},
\end{equation}
where $\sigma^\alpha \in \{X, Y, Z\}$ are the standard single-qubit Pauli matrices 
and $I$ is the $2\times 2$ identity. The resulting $L$-qubit operator 
$\hat{\mathcal{P}}^\alpha_i$ is a Pauli string - a tensor product of single-qubit 
Pauli operators over the full qubit register - with a non-trivial entry only at 
position $i$. Ground states sampled across the $(J_\perp/J_z, J'/J_z)$ parameter space are used to construct the QCNN training and test datasets.

\section{Results}
\label{sec:results}
In this section we apply the methods described above to the Fermi-Hubbard model with next-to-nearest neighbour hopping and the anisotropic Heisenberg model with next-to-nearest neighbour interaction, investigating the accuracies of QCNN training from ground states data, Pauli back propagation and training the QCNN model using classical surrogate simulation. 
\subsection{QCNN training}
\label{results:qcnn_training}

\begin{figure}[t!]
    \centering
    \includegraphics[width=\linewidth]{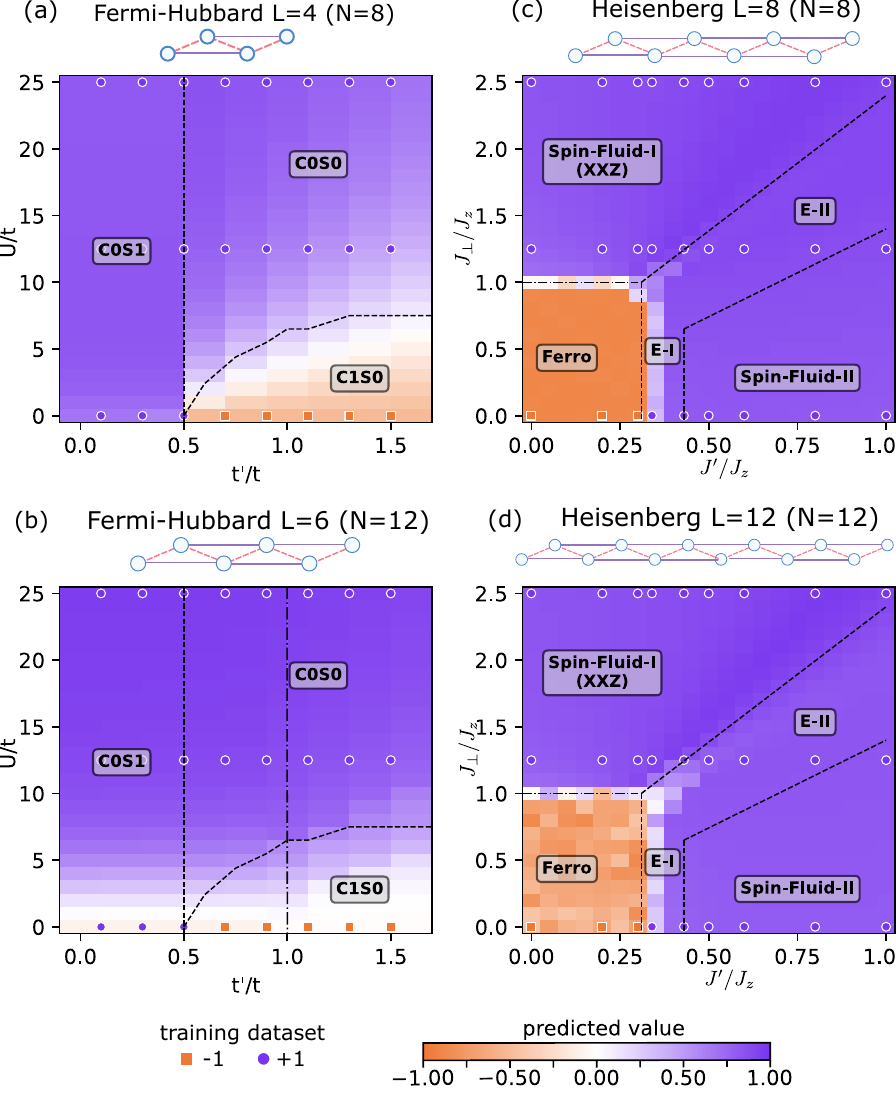}
    \caption{
    QCNN phase-differentiation results trained from $27$ labelled ground states. 
    Panels (a,b) show the extended Fermi-Hubbard model (metallic, marked with $-1$, versus non-metallic, marked with $+1$) for $N=8$ and $N=12$ qubits, respectively, and panels (c,d) show the extended Heisenberg model (ferromagnetic, marked with $-1$, versus non-ferromagnetic, marked with $+1$) for $N=8$ and $N=12$ qubits, respectively. Training points are marked by symbols (two classes), and dashed lines indicate approximate reference phase boundaries (thermodynamic-limit expectations), the dash-dotted line on panel (b) indicates the shift of C0S1-C0S0 boundary due to size effect. Heatmaps show the QCNN output evaluated across the full parameter grid after $200$ optimisation steps using the circuit family of Fig.~\ref{fig:circuit}. 
    }
    \label{fig:qcnn_many_training_dots_heatmaps}
\end{figure}

We first benchmark the baseline QCNN phase-differentiation pipeline described in Sec.~\ref{methods:qcnn} by training QCNNs directly on ground-state inputs for both the extended Fermi-Hubbard and anisotropic Heisenberg models. Unless stated otherwise, we consider chains of $L=4$ and $L=6$ Fermi-Hubbard sites (corresponding to $N=8$ and $N=12$ qubits after Jordan-Wigner encoding) and the analogous system sizes for the Heisenberg chain. For all results in this subsection, we simulate an ideal noiseless quantum processor via statevector simulation and optimise the circuit parameters using gradient descent with an adaptive learning rate (initial value $\eta_0=100$, increased by $10\%$ after a successful step and decreased by $33\%$ after a step that increases the loss). The QCNN circuits are implemented in a Python quantum machine-learning framework~\cite{Soni_QCNN_GitHub}, using custom circuit constructions that match the architectures defined in Sec.~\ref{methods:qcnn} and Appendix~\ref{appendix:structureN12}.

\paragraph{Data generation and labelling.}
For the Hubbard model, the training and test datasets are constructed from exact ground states obtained by exact diagonalisation at half filling (i.e., with the number of fermionic particles equal to the number of lattice sites, with no spin imbalance). To avoid numerical issues associated with ground-state degeneracies at exactly $U=0$, we replace $U/t=0$ by a small value $U/t=10^{-3}$ in the training set when needed (this affects only the datasets used for Fig.~\ref{fig:qcnn_many_training_dots_heatmaps}). For both models, each training point is assigned a binary label indicating membership in the target phase ($-1$) versus the complement ($+1$): metallic (MET) versus non-metallic phases for the Hubbard model, and ferromagnetic (FM) versus non-ferromagnetic phases for the Heisenberg model.

\paragraph{Training from moderately sparse labels.}
Figure~\ref{fig:qcnn_many_training_dots_heatmaps} shows QCNN training from a set of $27$ labelled points chosen to emulate the realistic situation where the ground state data and labels are only known for a set of few points. Notably, the training points are selected so that the phase boundary region is only sparsely sampled, meaning that a simple interpolation over parameters alone would not yield a reliable phase map. Despite this, the trained QCNNs recover qualitatively correct phase regions (indicated by dashed lines) in the case of Fig.~\ref{fig:qcnn_many_training_dots_heatmaps} (a),(c) and (d).
Fig.~\ref{fig:qcnn_many_training_dots_heatmaps}(a) shows the QCNN output for the extended Fermi-Hubbard model on an $L=4$ chain ($N=8$ qubits). The classifier assigns training values (depicted as small orange and purple dots in the phase diagrams) close to $+1$ throughout the C0S1 and C0S0 regions in the upper part of the diagram, while the metallic phase in the lower-right corner is assigned values close to $-1$. Between these two regions, the QCNN produces a relatively narrow crossover whose location follows the expected phase boundary (dashed line). Similarly, in Fig.~\ref{fig:qcnn_many_training_dots_heatmaps}(c) and (d), the QCNN correctly identifies the ferromagnetic region in the lower-left corner while assigning nearly uniform positive outputs throughout the XXZ and E-II regions. The transition follows the expected FM boundary (dashed) despite only a few training points being located in its vicinity.

In the case of Fig.~\ref{fig:qcnn_many_training_dots_heatmaps}(a), (c) and (d) the transition between the different regions occurs close to the reference phase boundaries (dashed lines) rather than midway between neighbouring training points, indicating that the circuit can extract phase-relevant features from the many-body ground-state inputs rather than from parameter proximity to labelled points (confusion matrices and additional diagnostics are provided in Appendix~\ref{appendix:confusion_matrices}). 

The main exception occurs in Fig.~\ref{fig:qcnn_many_training_dots_heatmaps}(b), \textit{i.e.}\ the $L=6$ ($N=12$ qubits) Hubbard case, where finite-size effects shift the relevant boundary compared to the thermodynamic-limit reference~\cite{bespalova2025simulating}. In particular, the transition around $t'/t\simeq 0.5$ expected in the bulk is substantially displaced in this specific system size, so a label assignment based solely on thermodynamic-limit expectations can become inconsistent with the finite-size spectrum used for training. In Appendix~\ref{appendix:confusion_matrices} we provide results obtained with a label assignment adapted to the finite-size behaviour, which restores consistent training performance.

\paragraph{Training with very few labelled points.}
To probe the limits of label efficiency, we further reduce the training set to only five labelled ground states (approximately $2$-$3$ per class). In contrast to the previous training scenarios, we also avoid collapsing multiple phases into a single class whenever possible, in order to test whether the QCNN output still carries meaningful structure beyond the imposed binary labelling.

\begin{figure}[t!]
    \centering
    \includegraphics[width=\linewidth]{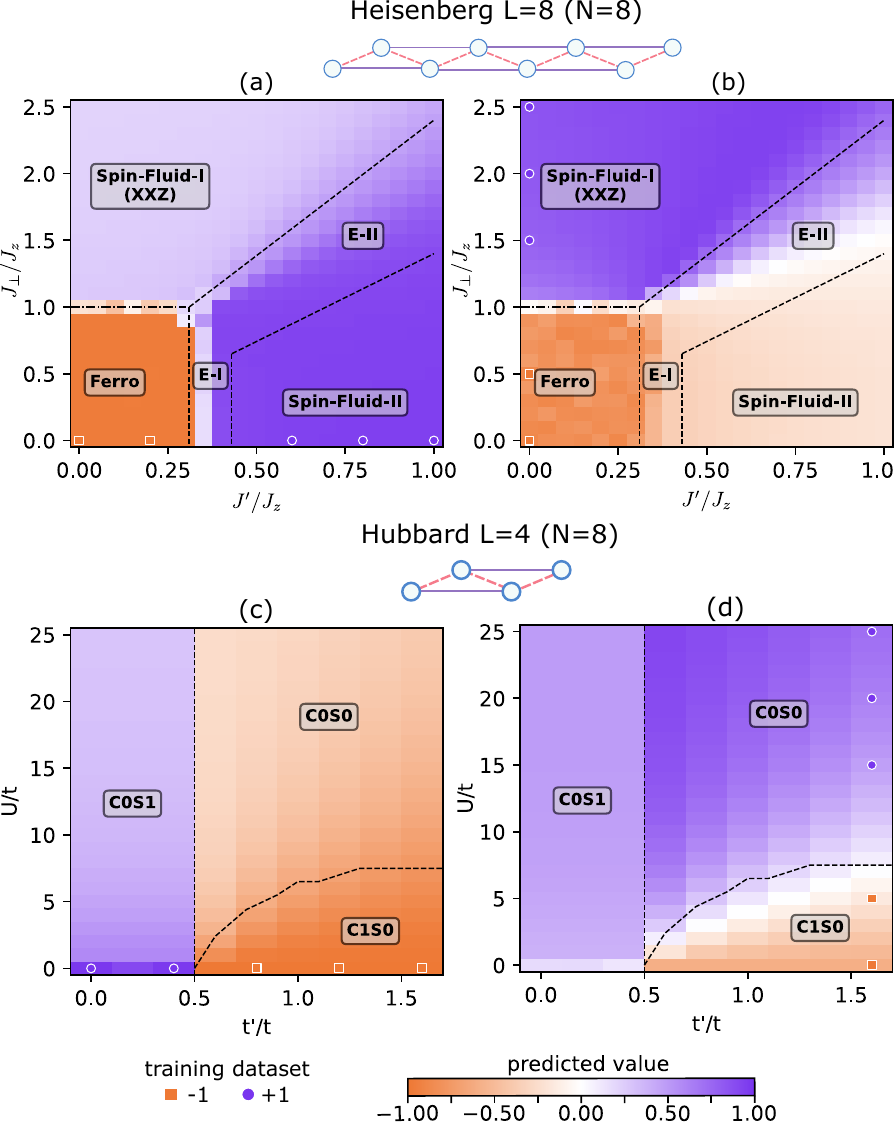}
    \caption{
    QCNN training from five labelled ground states. Heatmaps show the QCNN output across the full parameter grid after $200$ optimisation steps using the circuit family of Fig.~\ref{fig:circuit}. Training points are marked by symbols (two classes), and dashed lines indicate approximate reference phase boundaries. 
    Panels (a,b) correspond to Heisenberg training where the negative class ($-1$) is ferromagnetic and the positive class ($+1$) is chosen as either SF-II (a) or SF-I (b). Panels (c,d) correspond to Hubbard training, where the negative class ($-1$) is metallic C1S0 and the positive class ($+1$) is chosen as either spin-gapless C0S1 (c) or charge- and spin-gapped C0S0 (d).}
    \label{fig:qcnn_few_training_dots_heatmaps}
\end{figure}

Figure~\ref{fig:qcnn_few_training_dots_heatmaps}(a,b) demonstrates that, for the Heisenberg model, five labelled points can already be sufficient to recover the FM phase boundaries with good qualitative accuracy, even when the positive class is chosen as a specific competing phase (SF-I or SF-II). In both cases, the QCNN assigns values close to $-1$ throughout the ferromagnetic region, while the transition to positive values occurs close to the expected FM boundary (dashed line). In panel (a), where the positive class is chosen as SF-II, the SF-I and E-I regions acquire intermediate prediction values (close to $0$) instead of being sharply assigned to either class (so $\pm 1$). Likewise, in panel (b), where SF-I forms the positive class, the QCNN still distinguishes the E-I and XXZ regions through gradual variations of its output, even though these phases never appear in the training labels. The classifier does not simply extend the positive label uniformly into the remaining unlabelled phases. Instead, regions corresponding to phases that were not included in the binary labelling, such as E-I and XXZ, are assigned intermediate output values (close to $0$). This behaviour supports the conclusion that the QCNN learns features of the underlying many-body states rather than merely interpolating between the labelled points in parameter space. 

For the Hubbard model, training from five points is more sensitive to the placement and interpretation of labels. In Fig.~\ref{fig:qcnn_few_training_dots_heatmaps}(c), the chosen labels do not provide enough information to uniquely specify which physical distinction should be learned (e.g., spin-gapped versus spin-gapless versus charge-gapped versus charge-gapless), and the resulting decision landscape reflects this ambiguity. In Fig.~\ref{fig:qcnn_few_training_dots_heatmaps}(d), where the training points are placed deeper within representative regions of two phases, the QCNN output more clearly separates the corresponding regimes, although fine-grained phase boundaries remain difficult to infer from such a sparse label set.\\

The results above establish forward QCNN training as a baseline and provide trained parameter sets for subsequent analysis. In the following subsection we use these optimised QCNN parameters to backpropagate the final measurement observable and quantify how truncation affects (i) the fidelity of reproducing QCNN outputs and (ii) the complexity of the resulting Pauli expansion, which in turn determines the feasibility of classical-surrogate optimisation.

\subsection{Truncation accuracy and complexity of Pauli backpropagation}
\label{results:backprop_accuracy}

\begin{figure}[t]
    \centering
    \includegraphics[width=0.9\linewidth]{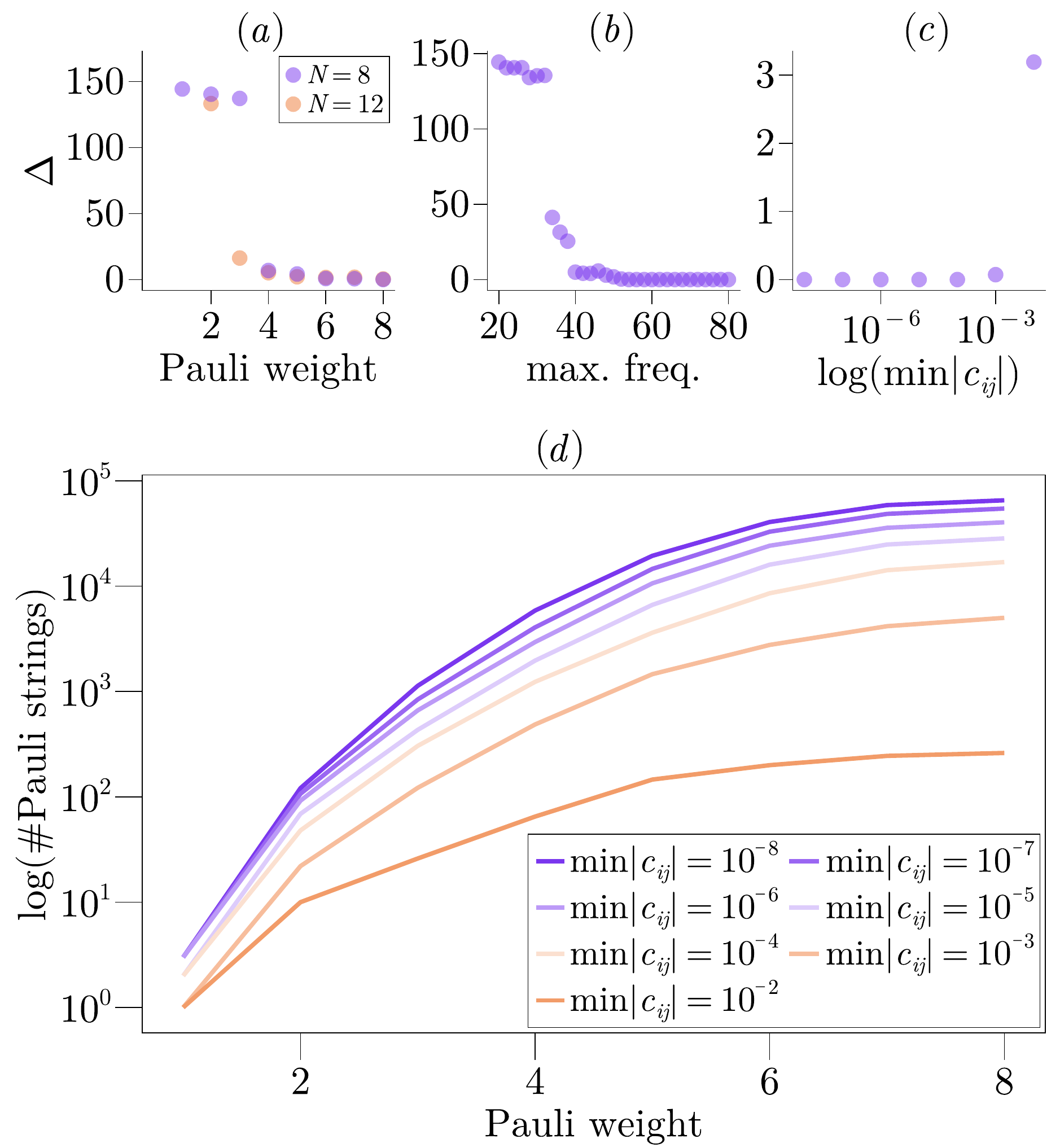}
    \caption{
Truncation accuracy and complexity of Pauli backpropagation for a trained QCNN. 
The final measurement observable (here $Z_7$ on the output qubit for both $N=8, 12$, see Fig.~\ref{fig:circuit},\ref{fig:QCNN_structure_N12}) is backpropagated through the optimised QCNN to obtain a truncated Pauli expansion, whose expectation value on each input state yields a backpropagated estimate $f_{\psi_n}^{\text{(PBP)}}(\boldsymbol{\theta_{\text{opt}}})$. 
Panels (a-c) report the endpoint deviation $\Delta$
between forward QCNN outputs $f_{\psi_n}(\boldsymbol{\theta_{\text{opt}}})$ and backpropagated estimates $f_{\psi_n}^{\text{(PBP)}}(\boldsymbol{\theta_{\text{opt}}})$ as a function of three truncation schemes: maximum Pauli weight (a), maximum frequency cutoff (b), and minimum absolute coefficient threshold (c), shown for $N=8$ (and $N=12$ in (a)) qubits. 
Panel (d) shows the corresponding number of retained Pauli strings in the truncated expansion (cost proxy) versus Pauli-weight cutoff for several coefficient thresholds for $N=8$.}
    \label{fig:backprop}
\end{figure}

\begin{figure}[t]
    \centering
    \includegraphics[width=1.0\linewidth]{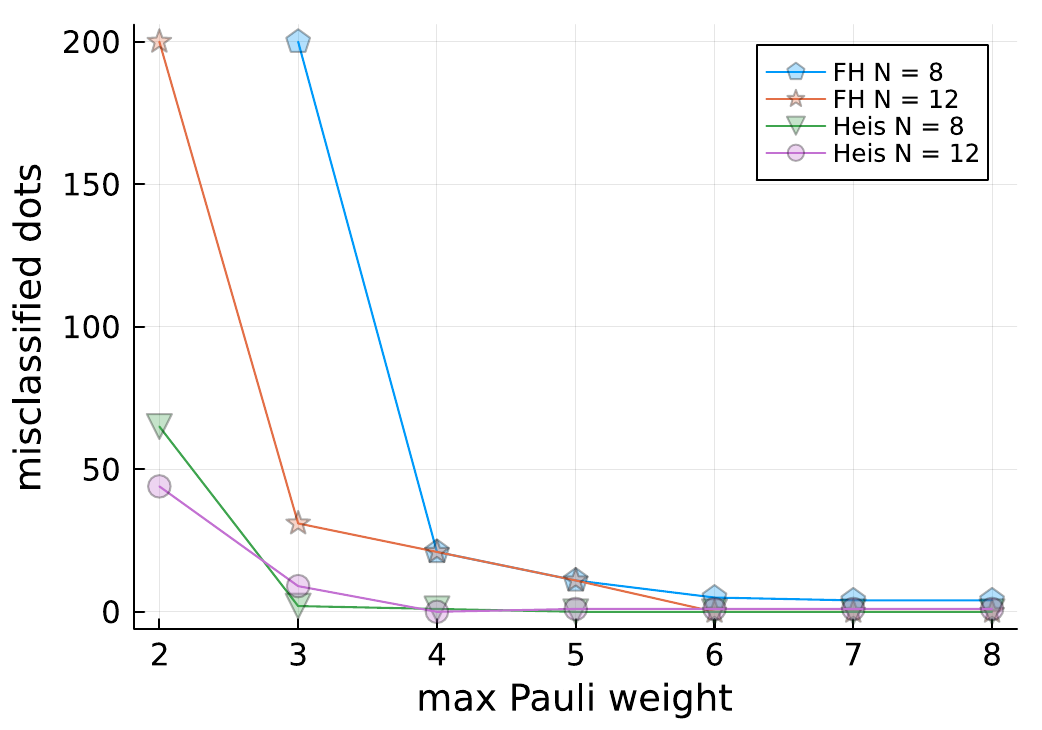}
\caption{
Truncation-induced classification disagreement as a function of Pauli-weight cutoff.
For each truncation setting, we assign labels using the same decision rule as in the forward QCNN evaluation (threshold on the prediction value) and count the number of phase-diagram points whose assigned class differs between the forward QCNN output and the truncated backpropagation estimate. Results are shown for $N=8$ and $N=12$ qubits.}
    \label{fig:misclassified}
\end{figure}

Since the classical-surrogate optimisation in Sec.~\ref{methods:fullyclassical} replaces circuit execution by truncated Pauli backpropagation, we first quantify (i) the approximation error induced by truncation and (ii) the associated complexity of the resulting Pauli expansion. Throughout this section we keep the QCNN parameters fixed to the optimised values obtained in the previous subsection and use Pauli backpropagation only as an evaluation primitive.

\paragraph{Backpropagated estimator.}
Starting from the final measurement observable on the output qubit, $\hat{O}=Z_7$, we backpropagate the observable through the trained QCNN using \texttt{PauliPropagation.jl}~\cite{rudolph2025pauli} (see Fig.~\ref{fig:backpropagation}). For each input point $n$ in the phase diagram and for each choice of truncation thresholds, Pauli backpropagation yields a truncated expansion
\begin{equation}
    f_{\psi_n}^{\text{(PBP)}}(\boldsymbol{\theta})  = \sum_i \alpha_i(\boldsymbol{\theta}) \,\langle P_i \rangle_{\psi_n}
\end{equation}
which provides an approximation to the direct QCNN prediction 
$ f_{\psi_n}(\boldsymbol{\theta}) $
obtained via forward circuit evaluation. To quantify the overall endpoint error introduced by truncation, we first consider the endpoint deviation
\begin{equation}
    \Delta = \sum_n \left(f_{\psi_n}(\boldsymbol{\theta_{\text{opt}}}) -f_{\psi_n}^{\text{(PBP)}}(\boldsymbol{\theta_{\text{opt}}})\right)^2,
\end{equation}
which measures the total difference between the forward QCNN outputs and their backpropagated approximations across the entire dataset. 
Results for $\Delta$ are summarised in Fig.~\ref{fig:backprop}.
While $\Delta$ provides a convenient global measure of approximation quality, it is not directly related to the performance of the phase classifier. Small deviations can change the predicted phase whenever they occur close to the decision boundary, whereas much larger deviations deep inside a phase may leave the assigned label unchanged. 
We therefore primarily quantify truncation-induced label disagreement: the number of input points 
whose predicted class changes when replacing $f_{\psi_n}(\boldsymbol{\theta_{\text{opt}}})$ by $f_{\psi_n}^{\text{(PBP)}}(\boldsymbol{\theta_{\text{opt}}})$. Results for the misclassifications are summarised in Fig.~\ref{fig:misclassified}.

\paragraph{Truncated endpoint deviation.}
Figure~\ref{fig:backprop} summarises the dependence of $\Delta$ on three common truncation schemes: a Pauli-weight cutoff, a minimum absolute coefficient threshold, and a maximum frequency cutoff. Overall, all three truncation schemes reproduce the expected behaviour: in Fig.~\ref{fig:backprop}(a), as more Pauli strings are retained, the endpoint deviation $\Delta$ decreases, which means the backpropagated prediction converges towards the exact QCNN output.  Restricting the maximum Pauli weight leads to most of the reduction in error already achieved by weights around five or six. 
The frequency cutoff in Fig.~\ref{fig:backprop}(b) exhibits a similar trend: allowing progressively higher maximum frequencies systematically reduces the endpoint deviation, with the improvement saturating once frequencies of approximately $50$--$60$ are included. In contrast, truncation based on a minimum absolute coefficient threshold (Fig.~\ref{fig:backprop}(c)) leaves the endpoint deviation essentially unchanged over several orders of magnitude in the threshold value, indicating that coefficients below this level contribute negligibly to the QCNN output. A noticeable increase in $\Delta$ is observed only when relatively large coefficients begin to be discarded. Together, these results suggest that the trained QCNN derives most of its predictive power from relatively low-weight Pauli operators. 

As expected, increasing the truncation budget (e.g., allowing larger Pauli weight or larger maximum frequency) systematically reduces $\Delta$ for both $N=8$ and $N=12$ qubits. Fig.~\ref{fig:backprop}(d) shows the corresponding growth in the number of retained Pauli strings, which we use as a proxy for classical cost and for measurement overhead in a prospective hardware implementation.
The number of terms grows rapidly, approximately exponentially with the allowed Pauli weight, and can differ by several orders of magnitude depending on the coefficient threshold. 

\paragraph{Truncation-induced misclassifications.}
As mentioned above, the decrease in $\Delta$ does not directly translate into classification performance.
We therefore also report the number of truncation-induced label changes in Fig.~\ref{fig:misclassified} as a function of the maximum Pauli weight. The disagreement count decreases with increasing weight, and we observe that a cutoff around $\mathrm{PW}\approx 6$ yields near-minimal disagreement for both $N=8$ and $N=12$ in the studied setting. Residual disagreements at large cutoff occur predominantly for points with predictions close to the decision boundary, where small numerical deviations can change the assigned label.\\

We emphasise that this section benchmarks endpoint fidelity of truncated backpropagation at a fixed, already-optimised parameter set. This is a necessary consistency check for using backpropagation as an evaluation primitive and for calibrating its cost. However, by itself it does not guarantee that an optimisation procedure using a truncated surrogate at every step will converge to the same optimum, since truncation can in principle distort the effective loss landscape away from $\boldsymbol{\theta}^\star$. We therefore assess trainability under truncation directly in Sec.~\ref{results:surrogate_training}.

\subsection{QCNN optimisation using classical surrogate simulation}
\label{results:surrogate_training}

\begin{figure}
    \centering
    \includegraphics[width=\linewidth]{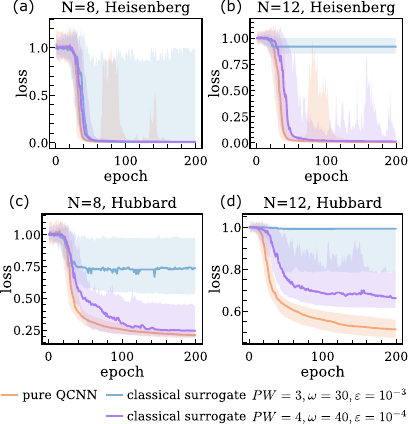}
    \caption{
Training-loss dynamics: forward QCNN training versus classical-surrogate optimisation based on truncated Pauli backpropagation.
Each panel shows the training loss versus epoch for (top row) the Heisenberg model and (bottom row) the Fermi-Hubbard model, using (left column) $N=8$ and (right column) $N=12$ qubits. 
Orange curves correspond to forward QCNN training, while blue/purple curves correspond to surrogate optimisation using truncated backpropagation with different truncation thresholds; the legend reports the truncation parameters and the associated cost proxy (number of retained Pauli strings in the backpropagated observable). 
Solid lines show the mean across $N_{\mathrm{seed}}$ random initialisations and shaded bands indicate the 10th-90th percentile interval. All runs use the same optimiser and learning-rate schedule as described in Sec.~\ref{methods:qcnn}.
}
    \label{fig:classical_surrogate_losses}
\end{figure}

The results of Sec.~\ref{results:backprop_accuracy} establish that truncated Pauli backpropagation can reproduce the outputs of a trained QCNN at a fixed parameter point, and it quantifies the growth in Pauli-list size as truncation is relaxed. However, surrogate trainability is a stronger requirement: the truncated surrogate must provide a sufficiently accurate learning signal throughout the optimisation trajectory. We therefore perform end-to-end optimisation runs in which the QCNN forward evaluation is replaced by the truncated backpropagation surrogate at every optimisation step, as described in Sec.~\ref{methods:fullyclassical}.

\paragraph{Setup and comparison protocol.}
We implement the surrogate training loop using a Julia-based Pauli backpropagation library~\cite{rudolph2025pauli} and match the QCNN training protocol of Sec.~\ref{methods:qcnn} as closely as possible: we use the same circuit architecture at $N=8$ and $N=12$ qubits, the same optimiser and learning-rate schedule, and the same training/test splits. The only difference is the evaluation of model outputs: in the QCNN case they are obtained by forward execution of the QCNN circuit, whereas in the surrogate case they are computed from a truncated Pauli expansion of the backpropagated observable.

\paragraph{Training dynamics under truncation.}
Figure~\ref{fig:classical_surrogate_losses} compares the training loss as a function of epoch for direct QCNN training and for surrogate training under several truncation regimes. 
For the Heisenberg model $N=8$ in Fig.~\ref{fig:classical_surrogate_losses}(a), we find that surrogate training converges reliably already under aggressive truncations ($\text{PW}=3$, $\omega=30$, $\varepsilon=10^{-3}$ and $\text{PW}=4$, $\omega=40$, $\varepsilon=10^{-4}$)   
in the studied setting, and the resulting loss dynamics closely track the QCNN baseline. For $N=12$ (Fig.~\ref{fig:classical_surrogate_losses}(b)), this behaviour remains true for ($\text{PW}=4$, $\omega=40$, $\varepsilon=10^{-4}$), whereas the truncation ($\text{PW}=3$, $\omega=30$, $\varepsilon=10^{-3}$) leads to an almost constant loss around $\lambda = 0.9$ throughout training. 
Thus, while surrogate training remains effective for the Heisenberg model, the acceptable truncation becomes more restrictive as the system size increases. In contrast, for the Fermi-Hubbard model in Fig.~\ref{fig:classical_surrogate_losses}(c) and (d) exhibits a much stronger dependence on the truncation level. Here, the same aggressive truncation leads to early saturation at a significantly higher loss, particularly for $N=12$. Surrogate runs either plateau at higher loss or exhibit substantially increased run-to-run variability. Achieving consistent convergence requires relaxing the truncation thresholds, in particular increasing the allowed Pauli weight, which significantly increases the Pauli-list size and thus the evaluation cost.

A plausible origin of this disparity is the effective non-locality introduced by fermion-to-qubit mappings such as the Jordan-Wigner transformation: even when the underlying fermionic operators are local, their qubit-representations can involve long Pauli strings, which can amplify the truncation burden for backpropagated observables. While a detailed operator-level interpretation is beyond the scope of this work, the observed optimisation behaviour directly demonstrates that surrogate trainability is strongly model-dependent and cannot be inferred from endpoint fidelity alone.

\paragraph{Cost implications (Pauli-list size and measurement overhead)}
The main source of costs in the classical surrogate procedure on quantum data comes from measuring the expectation values of required Pauli strings. In a shot-based implementation, estimating each term to additive precision $\epsilon$ requires $O(1/\epsilon^2)$ samples and thus $O(N_\text{Pauli strings}\epsilon^{-2})$ samples when the expectation values of $N_\text{Pauli strings}$ Pauli strings are required. This amount can be reduced to $O(4^\text{PW} \log(N_\text{Pauli strings})\epsilon^{-2})$ with the use of classical shadow techniques~\cite{huang2020predicting}.
We summarise in Table~\ref{tab:costs} the average numbers of Pauli terms in the truncated backpropagated observables used in classical surrogate procedure, with the ``total'' column corresponding to the maximum number of terms with the set Pauli weight PW, and ``mean on $\boldsymbol{\theta}_\text{rand/opt}$'' columns corresponding to the average number of terms per optimisation step in the beginning of the procedure when the parameters are randomised and in the end of the procedure when they are close to optimal respectively. The two latter depend on the type of circuit
structure we use, while the last one depends also on the optimisation problem and is averaged in our case across the Fermi-Hubbard and Heisenberg model runs. 
We note that except for the cases when a very small number of iterations are required or the training initially starts from the parameters already close to an optimal point, it is generally beneficial to first precompute the expectation values of all the possible PW-local Pauli strings (scaling as $ O\left( \frac{N! 4^\text{PW}}{(N-\text{PW})!\text{PW}! }\right)$) as the optimisation steps have large overlap in the strings used. Alternatively, one may cache the expectation values as they appear and only compute the missing ones in subsequent steps. Another important observation is that the minimal absolute coefficient value $\varepsilon$ required in the training generally has the same magnitude as the required precision $\epsilon$ of the expectation value, both contributing to the accuracy of Pauli terms in the sum. We thus can estimate the total number of shots required for the training procedure in the system of $N$ qubits, truncated with Pauli weight PW and minimal absolute coefficient value $\varepsilon$ as $O\left( 4^\text{PW} \log{\left(\frac{N! 4^\text{PW}}{(N-\text{PW})!\text{PW}! }\right)} \varepsilon^{-2}\right)$.

\begin{table}[tb]
\centering
\renewcommand{\arraystretch}{1.4}
\begin{tabular}{|c|c|c|c|c|c|c|}
\hline
\multirow{2}{*}{PW} &
\multirow{2}{*}{$\omega$} &
\multirow{2}{*}{$\varepsilon$} &
\multirow{2}{*}{$N$} &
\multicolumn{3}{c|}{$N_\text{Pauli strings}$} \\ \cline{5-7}
 &  &  &  & total & mean on $\boldsymbol{\theta}_\text{rand}$  & mean on $\boldsymbol{\theta}_\text{opt}$\\ \hline
\multirow{2}{*}{3} & \multirow{2}{*}{30} & \multirow{2}{*}{$10^{-3}$} & 8  & $1789$  & $135 \pm 29$ & $41\pm 40$ \\ \cline{4-4}\cline{5-7}
                   &                    &                            & 12 &  $ 6571 $ & $15\pm 9$ & $9\pm 8$ \\ \hline
\multirow{2}{*}{4} & \multirow{2}{*}{40} & \multirow{2}{*}{$10^{-4}$} &8  &  $7459$ & $2233\pm 154$ & $1081 \pm 493$ \\ \cline{4-4}\cline{5-7}
                 &                    &                            & 12 &   $46666 $ & $1149\pm 291$ & $382 \pm 240$ \\ \hline
8 & \multirow{2}{*}{$100$} & \multirow{2}{*}{$10^{-8}$} & 8  & $65536 $ & $65535\pm 2$ & $65448\pm 276$ \\ \cline{1-1}\cline{4-4}\cline{5-7}
12                   &                      &                       & 12 &   $16777216$ & $1.67757(24) \cdot 10^7$ & $1.644(66) \cdot 10^7$ \\ 
\hline
\end{tabular}
\caption{Summary of amounts of Pauli strings needed for Pauli back propagation in different truncation regimes. Column "total" provides the overall number of Pauli strings of length $N$ and up to PW non-identity entries; column "mean on $\boldsymbol{\theta}_\text{rand}$" provides the average amount ($\pm$ standard deviation) of Pauli strings in back propagation of a circuit described in \ref{methods:qcnn} with random parameters (averaged over $100$ sets of random parameters); column "mean on $\boldsymbol{\theta}_\text{opt}$" provides the same values but for $100$ sets of parameters that resulted from optimisation procedures presented in Fig.~\ref{fig:classical_surrogate_losses},
trained with corresponding PW, $\omega$, $\varepsilon$ parameters (in case of the last two rows we take optimal parameters from pure QCNN training due to high costs of classical surrogate optimisation in this regime), results are averaged over both Fermi-Hubbard and Heisenberg models. }
\label{tab:costs}
\end{table}

\section{Discussion and Conclusion}
\label{sec:conclusion}

In this work, we evaluated the extent to which truncated Pauli backpropagation surrogates can replace QCNN evaluation and training for ground-state phase differentiation of strongly correlated one-dimensional many-body systems. Three results emerge. First, trained fully quantum QCNN models achieve reliable phase differentiation for both spin and fermionic models within the investigated finite-size settings. Second, truncated Pauli backpropagation exhibits a clear accuracy–complexity trade-off, which is clearly assessed through label-disagreement count. Third, and most important, endpoint accuracy does not by itself guarantee surrogate trainability, which we instead found to be strongly dependent on the target class of physical systems: surrogate training converges robustly under strong truncation for the Heisenberg model, whereas the fermionic Hubbard model requires substantially lower truncation thresholds. This results most likely from both the stronger non-local nature of genuine fermionic correlations and the effect of the Jordan-Wigner fermion to qubit mapping we employed.

This model dependence immediately motivates a deeper, natively fermionic treatment of the surrogate, which we regard as the most promising direction for future work. For instance, more compact or locality-preserving encodings~\cite{o2024ultrafast, havlivcek2017operator} could keep the weight of backpropagated observables under control even for fermionic models, directly reducing the truncation burden that limits Hubbard-model trainability here. A complementary route is to bypass the qubit picture altogether and propagate directly in the Majorana representation~\cite{miller2025simulation}, so that the surrogate expansion is organised by fermionic rather than Pauli locality and is thus less sensitive to qubit-space non-locality. Consistently, the measurement primitive could shift from Pauli-based classical shadows to fermionic Gaussian shadows~\cite{zhao2021fermionic}. Beyond encodings, the systematic study of scaling with system size, circuit depth, and training-set composition, the replacement of exact-diagonalisation inputs by explicit state-preparation procedures with their associated errors, and a physical interpretation of the dominant backpropagated terms in terms of correlators or order parameters would all represent natural next steps.

Taken together, our findings offer a promising blueprint for the design of specialised hybrid workflows, powered by classical high performance tools and shadow measurements, to characterise and classify quantum states produced in typical quantum simulation experiments on both analog and digital architectures.

\section{Acknowledgments}
This research has received funding from the European
Union’s Horizon 2020 research and innovation programme
under the Marie Skłodowska-Curie Grant Agreement No.
955479 (MOQS - Molecular Quantum Simulations). This research was supported by the
project RESQUE (Rethinking Quantum Simulations in
the Quantum Utility Era, grant number 20QU-1 225229)
and by the National Center of Competence in Research
(NCCR) SPIN (grant number 225153), both funded by
the Swiss National Science Foundation. T.~B.~and J.~S.~have been supported by the ERC Consolidator project MATHLOCCA (Grant nr.~101170485). R.~S.~is supported by the MLQ program that has received funding from the European Union’s Horizon Europe research and innovation programme under the Marie Skłodowska-Curie grant agreement number 101120240.
This work extends preliminary results on QCNN-based Fermi-Hubbard phase classification obtained during L. M.'s Master's thesis, supported by the Inspire Potentials -- NCCR Marvel Master’s Fellowship.
The authors used Claude (Anthropic; Claude Sonnet 4.6) for language editing (grammar and style). 

\bibliography{bibliography}

\appendix

\section{QCNN structures and gate definitions}
\label{appendix:structureN12}
\begin{figure}[tb]
    \centering
    \includegraphics[width=\linewidth]{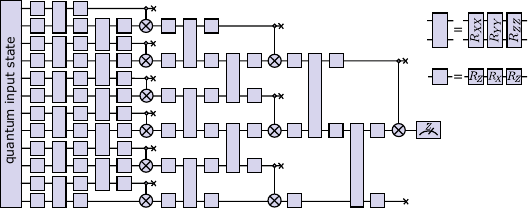}
    \caption{QCNN structure for $N=12$ qubits used in this work. Each parametrised gate is parametrised independently.}
    \label{fig:QCNN_structure_N12}
\end{figure}

Here in Fig.~\ref{fig:QCNN_structure_N12} we present the full structure of our QCNN circuit in the $N=12$ case (for both Heisenberg and Fermi-Hubbard models), which is a natural extension of the circuit presented in Figure~\ref{fig:circuit} after reducing redundant gates.

For any $N$, each of rotation gate $R_X(\theta)$, $R_Y(\theta)$, $R_Z(\theta)$, $R_{XX}(\theta)$, $R_{YY}(\theta)$, $R_{ZZ}(\theta)$ depends on its own independent parameter. The gates are defined as follows: 

\[
R_X(\theta)=e^{-i\frac{\theta}{2}X}
=
\begin{pmatrix}
\cos\!\left(\frac{\theta}{2}\right) & -i\sin\!\left(\frac{\theta}{2}\right)\\
-i\sin\!\left(\frac{\theta}{2}\right) & \cos\!\left(\frac{\theta}{2}\right)
\end{pmatrix},
\]

\[
R_Y(\theta)=e^{-i\frac{\theta}{2}Y}
=
\begin{pmatrix}
\cos\!\left(\frac{\theta}{2}\right) & -\sin\!\left(\frac{\theta}{2}\right)\\
\sin\!\left(\frac{\theta}{2}\right) & \cos\!\left(\frac{\theta}{2}\right)
\end{pmatrix},
\]

\[
R_Z(\theta)=e^{-i\frac{\theta}{2}Z}
=
\begin{pmatrix}
e^{-i\frac{\theta}{2}} & 0\\
0 & e^{i\frac{\theta}{2}}
\end{pmatrix},
\]

\begin{align*}
    R_{XX}(\theta)= & e^{-i\frac{\theta}{2}(X\otimes X)}
=\\
&\begin{pmatrix}
\cos\!\left(\frac{\theta}{2}\right) & 0 & 0 & -i\sin\!\left(\frac{\theta}{2}\right)\\
0 & \cos\!\left(\frac{\theta}{2}\right) & -i\sin\!\left(\frac{\theta}{2}\right) & 0\\
0 & -i\sin\!\left(\frac{\theta}{2}\right) & \cos\!\left(\frac{\theta}{2}\right) & 0\\
-i\sin\!\left(\frac{\theta}{2}\right) & 0 & 0 & \cos\!\left(\frac{\theta}{2}\right)
\end{pmatrix},
\end{align*}

\begin{align*}
R_{YY}(\theta)= &
e^{-i\frac{\theta}{2}(Y\otimes Y)}
=\\
&
\begin{pmatrix}
\cos\!\left(\frac{\theta}{2}\right) & 0 & 0 & i\sin\!\left(\frac{\theta}{2}\right)\\
0 & \cos\!\left(\frac{\theta}{2}\right) & -i\sin\!\left(\frac{\theta}{2}\right) & 0\\
0 & -i\sin\!\left(\frac{\theta}{2}\right) & \cos\!\left(\frac{\theta}{2}\right) & 0\\
i\sin\!\left(\frac{\theta}{2}\right) & 0 & 0 & \cos\!\left(\frac{\theta}{2}\right)
\end{pmatrix},
\end{align*}

\[
R_{ZZ}(\theta)=e^{-i\frac{\theta}{2}(Z\otimes Z)}
=
\begin{pmatrix}
e^{-i\frac{\theta}{2}} & 0 & 0 & 0\\
0 & e^{i\frac{\theta}{2}} & 0 & 0\\
0 & 0 & e^{i\frac{\theta}{2}} & 0\\
0 & 0 & 0 & e^{-i\frac{\theta}{2}}
\end{pmatrix}.
\]

In the situation where local Pauli observables are effectively enough, the quantum version of QCNN training can also be simplified. The simplification of the circuit leads to reduce of the number of parameters in the optimisation and significantly eases the optimal point search. One of the approaches for such a simplification is to choose a circuit that would reflect the locality of effective observables.

\begin{figure}[tb]
    \centering
    \includegraphics[width=\linewidth]{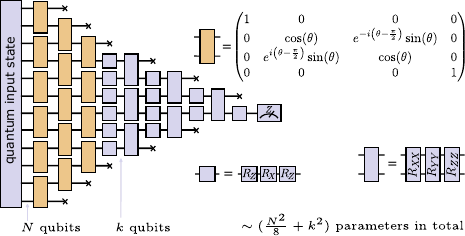}
    \caption{Example of simplifyed QCNN circuit that effectively implements the structure with reduced number of parameters, accommodating for the sufficiency of k-local Pauli strings in the backpropagated expansion. }
    \label{fig:local_struct}
\end{figure}

An example of such a circuit is provided in Fig.~\ref{fig:local_struct}. There the right (gray) part of the circuit corresponds in the Heisenberg picture to creating the parametrised combination of all possible Pauli strings of length $k$. The left (orange) part implements effectively the scattering of of those $k$ Paulis across all the $N$ qubits. As it is impossible to have a two-qubit gate that would be doing a smooth transition between Identity and SWAP gate while conserving the locality of Pauli strings through a continuous range of parameter values, we suggest the second-best option: using the gate that is Identity at $\theta = 0$ and SWAP at $\theta = \frac{\pi}{2}$. This structure only requires $\sim \left(\frac{N^2}{8} + k^2\right)$ parameters, while the general one presented in Fig.~\ref{fig:circuit} requires $\sim \left(\frac{9}{8} N^2\right)$ parameters thus reducing the search space by up to $88\%$ in case of constant $k$.

\begin{figure}[h!]
    \centering
    \includegraphics[width=0.75\linewidth]{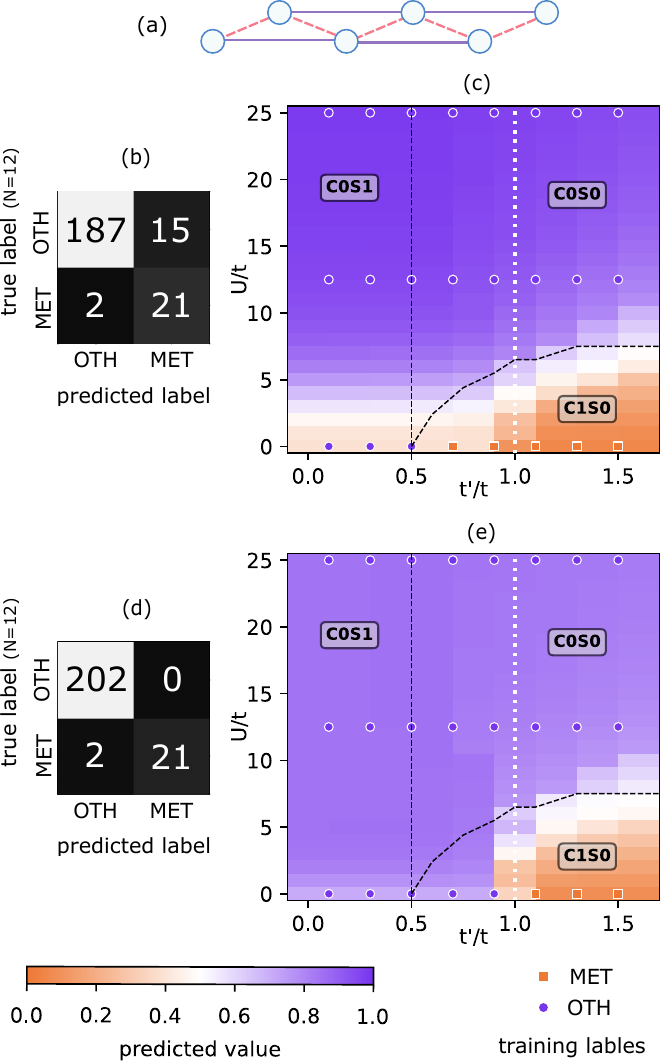}
    \caption{Phase diagrams computed from the trained QCNN for $N=12$ qubits, gradient descent with $500$ iterations. With training labels phase transition at $t'/t=0.5$ (top plot) and $t'/t=1.0$ (bottom plot). Round dots represent training data.  The heat map depicts the result of the gradient descent parameter optimisation with the QCNN structure shown in Appendix~\ref{appendix:structureN12}. The black dashed line refers to the approximate metallic-insulator phase boundary in the thermodynamic limit.}
    \label{fig:qcnn_12q_heatmaps}
\end{figure}
\section{$N=12$ Fermi-Hubbard case}
\label{appendix:confusion_matrices}

This section presents the resulting heatmaps that did not make it into the main text, plus the confusion matrices for some of them. 

In Fig.~\ref{fig:qcnn_12q_heatmaps}, we illustrate how a size effect can affect the simulation. For $6$-Hubbard-sites model, the natural change of structure of the ground state happens at $t'=1.0t$ instead of $t'=0.5t$ (as explained for the closed boundary condition in~\cite{bespalova2025simulating}). Thus, when we label the train data in a manner that does not respect the inner structure of the system and try to train the QCNN on this data, we receive the phase boundary qualitatively different from the physically relevant one. This behaviour signifies that to obtain the phase boundary in the thermodynamic limit, we need to increase the system size on which we train the model. 

\end{document}